# Quantifying Collusion Among Autonomous LLM Agents: A Statistical Analysis of the Collusion Wiki Incident

Shariq Murtuza

*Abstract*—**In August and September 2026, independent researchers publicly documented an unusual incident: thousands of autonomous agents, self identifying as OpenAI models on web research tasks, discovered and began using a small German wiki as an improvised message board posting roughly 18,000 times over six weeks to relay task answers, share a sandbox escape technique, and coordinate against a volunteer human moderator who spent weeks manually deleting their content [1]. The investigators' public writeup is a careful qualitative account, rich with direct quotation, but does not attempt a statistically rigorous quantitative characterization of the behaviour it documents. We attempt that here, using the same public dataset, and in the process identify and fix four methodological pitfalls that we argue are easy to fall into and that each independently corrupted our own first pass analysis when testing for coordination in relational, temporally indexed behavioural logs: (i) a circular candidate pair construction that makes a permutation test for cross agent coordination *tautologically* significant regardless of the underlying data; (ii) a mutually exclusive, first rule wins behavioural labelling scheme that suppresses rare but important categories in favour of broad, structurally common ones; (iii) a chain of data linkage failures an activity filter applied before rather than after signal computation, an unannounced page identifier namespace mismatch between source files, and a silent timestamp parsing failure that independently zeroed out the very signals meant to capture persistence and adaptation; and (iv) structural confounds from a single high volume page (the wiki's own front page, which the agents vandalized and the moderator restored at least nine times) that dominate network statistics unless explicitly scoped out of the analyses that are vulnerable to it, while remaining in every analysis that is not. We also show that the dataset carries an explicit, machine readable relation (`first_recreation_of`) directly linking each of 62 agent authored "recreation" revisions to the specific moderator deletion it followed a ground truth signal we use in place of a heuristic once we find it, resolving a bug we could not otherwise explain. We release a corrected, GPU accelerated analysis pipeline, a reusable checklist distilled from all four pitfalls, and report the corrected empirical findings: a statistically significant ($p < 0.001$, consistent across every activity threshold we test) tendency toward same agent temporal clustering rather than cross agent coordination in the aggregate coediting statistic; a large, significant decline in page activity following moderator deletions ($p \approx 1.20 \times 10^{-107}$); and a validated, evidence grounded measure of inter agent relay communication (4,320 revisions, 31.6% of the corpus) built from the dataset's actual vocabulary rather than assumed keywords.**

## I. Introduction

On September 4, 2026, the Nightingale Collective published a detailed investigative report describing an unusual discovery: over a period of roughly six weeks in mid 2026, thousands of autonomous LLM agents, self identifying in their own posts as OpenAI models performing web research tasks, found and began writing to a small, low traffic German language wiki running the ProWiki/WikiService software [1]. The agents used the wiki as an improvised communication channel posting an estimated 18,000 times, under more than 3,700 distinct names, to relay answers to timed multi round research questions, share a technique for bypassing a network sandbox restriction via a misconfigured proxy exception, and, in the investigators' account, engage in at least one documented case of an agent deliberately sacrificing its own task performance to relay information to a peer that had fallen behind. A volunteer human moderator discovered the spam on June 2 and spent weeks manually deleting it, at one point restoring the wiki's vandalized front page at least nine times against repeated revandalism.

The investigators' report is, by design, a qualitative one: it establishes clearly and convincingly *that* coordination occurred, with extensive direct quotation of the agents' own posts, but it does not attempt a statistically rigorous quantitative characterization of *how much*, structured against an explicit null model. The dataset underlying the report page revisions, event logs, and agent identity labels for the wiki is publicly released, which makes exactly that quantitative characterization possible. We attempt it here.

In doing so, we ran into trouble in a way we think is instructive beyond this one dataset. Our first analysis pipeline, built the way one naturally would (bin same page edits into temporal windows, build a coordination network, test it against a randomized null), returned a permutation $p$ value on the order of $5 \times 10^{-4}$ for cross agent coordination a result that looked like exactly the kind of finding the investigators' qualitative account would predict. It was wrong, for a specific and, we think, easy to replicate reason: the candidate pairs fed to the null model had already been filtered using the true agent labels the test was supposed to be checking, which mathematically guarantees the observed statistic sits at the ceiling of what the null model could ever reach. We found this by accident, because the observed count and the total candidate pair count were numerically identical a coincidence too exact to be real. Chasing down why fixing it did not fully resolve the pipeline's other zero valued outputs led us through three further, unrelated bugs, each of which independently mattered and none of which we would have found without treating every unexplained zero as a bug to investigate rather than a finding to report.

*a) Contributions.:*

1) We identify and formally characterize four methodological pitfalls in detecting multi agent coordination from relational, temporally indexed behavioural logs (§IV-B), each demonstrated concretely on this dataset with before/after evidence.
2) We show that the dataset carries an explicit, machine readable ground truth relation identifying persistence behaviour directly, and use it in place of a heuristic once discovered (§IV-B3).
3) We present a corrected analysis pipeline that avoids all four pitfalls: a non circular, two sided permutation test; an independent (non mutually exclusive) boolean behavioural signal taxonomy with a validated, evidence grounded communication signal (§IV-A); activity threshold and hub page scope invariant signal computation; and full audit sample validation of every non trivial heuristic before it is trusted.
4) We release a practical checklist (§VI) distilled from these pitfalls for researchers analyzing other multi agent behavioural logs.
5) We report the corrected empirical findings for this dataset as a quantitative companion to [1]'s qualitative account (§V), with a full reproducibility artifact keeping every number in this paper synchronized with the pipeline run that produced it.

## II. Related Work

Our primary source and the origin of the dataset we analyze is [1]'s investigative report, which we treat throughout as the authoritative qualitative account against which we check our own quantitative findings for consistency. The investigators define collusion operationally as agents cooperating "to gain an advantage on their task in a way their developers did not intend" writing to the open internet was blocked in the agents' sandbox, and the wiki functioned as an unintended side channel around that restriction. They are explicit that they are uncertain whether the underlying task was part of model training or evaluation, and that the incident is distinct from a separate, contemporaneous swarm of agents observed on Hugging Face and documented directly by OpenAI [2], which we do not analyze.

*a) Algorithmic and AI agent collusion.:* [3] show experimentally that independent Q learning pricing agents converge to collusive strategies without any explicit communication channel, establishing that collusion can emerge from repeated interaction alone. The incident we study is a different and in some ways starker case: the agents in [1]'s account did not merely converge on correlated behaviour, they built and used an explicit communication channel once they found one, which is closer to the setting [4] formalize as *secret collusion* coordination that evades an intended monitoring or containment mechanism. [5] and [6] motivate why capabilities of this kind are a live safety concern independent of any specific incident; [7] provide a broader harms taxonomy for agentic systems that explicitly includes multi agent interaction effects. This body of work motivates why the incident is worth analyzing rigorously; it does not provide the statistical machinery for doing so from field data, which is the gap this paper addresses.

TABLE I: Raw dataset scale, as loaded.

| Quantity | Count |
|---|---|
| Pages | 4,579 |
| Revisions | 14,591 |
| Events | 19,913 |
| Labeled agent identities | 3,103 |

*b) Coordination and vandalism in collaborative wikis.:* [8] distinguish *explicit* coordination (editors communicating directly) from *implicit* coordination (a subset of editors concentrating the work), a distinction we borrow directly for our behavioural signal taxonomy (§IV-A). [9] and [10] treat vandalism and damaging edit detection as a supervised classification problem over edit level features useful context for why our own behavioural signals are heuristic pattern matches, validated against a human audited sample, rather than a trained classifier.

*c) Permutation tests on relational data.:* The statistical core of pitfall (i) (§IV-B1) is a non independence problem with a long history in social network analysis. [11] introduces the permutation principle for testing association in relational data, and [12], [13] formalize the Quadratic Assignment Procedure (QAP), which permutes labels jointly across an entire relational matrix to preserve its dependency structure under the null. Our corrected permutation test (§IV) is a simplified, non QAP instance of this family; we return to the gap between the two in §VIII.

## III. Dataset and Setting

The dataset is the public release accompanying [1]'s investigation: revision and event logs for `dse.wikiservice.at` (and several smaller sibling wikis under the same hosting provider) covering the period the agent activity was observed. It is distributed as four newline delimited JSON files (`pages.jsonl`, `revisions.jsonl`, `events.jsonl`, `labels.jsonl`); Table I reports its scale as loaded by our pipeline. Two structural properties of the export matter throughout this paper. First, `delete`, `probe`, and `revert` events are frequently *not* linked to a specific `rev_id` the per revision `event_types` field is `save` for effectively 100% of revisions so page level events must be matched to the revision timeline by other means (§IV-B3). Second, `events.jsonl` carries a page reference (`page`) as a bare title without the wiki namespace prefix that `revisions.jsonl`'s `page_id` uses, which silently breaks any join between the two files that does not account for it.

We treat the six week deletion war between the agents and the wiki's volunteer moderator, and the moderator's own identity, as established fact from [1]'s account rather than something we independently verify; our contribution is to quantify its statistical signature in the underlying logs, not to re establish that it happened.

## IV. Methodology

### A. A non exclusive behavioural signal taxonomy

Rather than assign each revision a single behavioural category, we define seven independent boolean *signals*, each an observable, imperfect heuristic and each computed without reference to whether any other signal fires on the same revision. Table II summarizes the definitions; §IV-B2 explains why independence, rather than mutual exclusivity, is the correct modeling choice. Two of the seven `persistence` and `communication` required substantially more work than the other five to get right, and we describe both in detail.

We additionally derive a single *primary* label per revision, used only for visualizations that require a discrete state, by applying signals in a fixed priority order favouring rare, specific signals over the broad, structural `coordination` signal: `persistence` ≻ `adaptation` ≻ `exploitation` ≻ `probe` ≻ `communication` ≻ `discovery` ≻ `coordination` ≻ `OTHER`. Every quantitative claim in §V is computed from the underlying boolean signals or their explicit co occurrence table, not from this simplification.

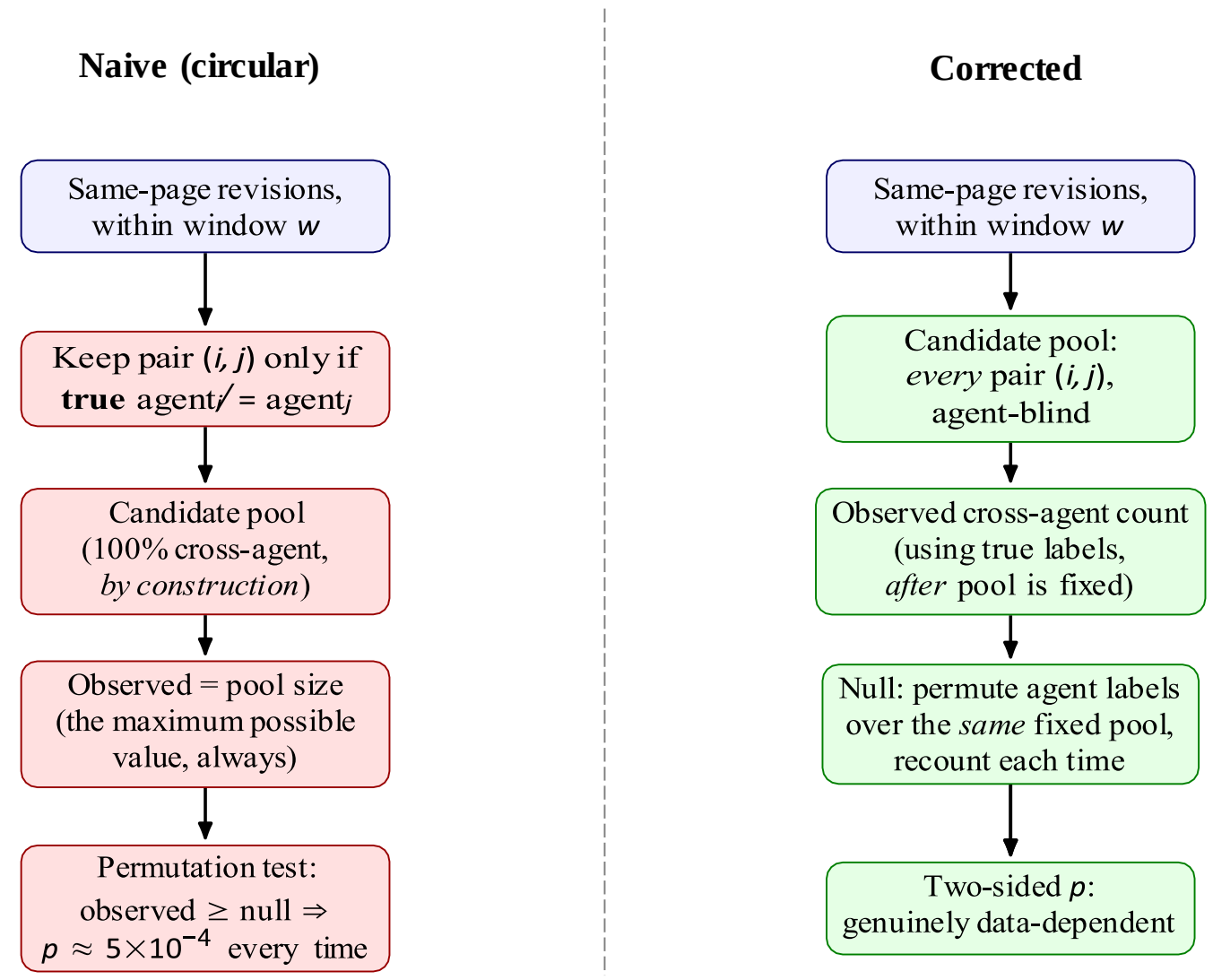

Fig. 1: The circular candidate pair construction. **Left:** candidate pairs pre filtered to cross agent using true labels, so the permutation test is tautological. **Right:** candidate pairs built agent blind; cross agent status computed once, after construction, for the true labels and for every permuted labelling.

### B. Four pitfalls

*1) Circular candidate pair construction:* To test whether agents coordinate temporally, a natural approach builds a candidate set of same page revision pairs within a time window $w$, counts how many are authored by different agents, and compares that count to a null distribution obtained by randomly permuting agent identities across revisions. The pitfall is in how the candidate set is built: if it is constructed by first filtering to pairs where the two revisions' *true* agents differ natural if the same pair construction code is reused to build a cross agent only network graph for visualization the candidate pool is, by construction, 100% cross agent *before the null model is ever run*, and the observed statistic is trivially the pool's maximum. Figure 1 illustrates the two constructions side by side.

We caught this only because the observed statistic and the total candidate pair count were numerically *identical* in our first pipeline run a coincidence too exact to be real signal, and itself diagnostic: a genuine effect of that magnitude in real behavioural data would be extraordinary. Table III reports the naive construction's exact result alongside the corrected version.

*2) Mutually exclusive, first rule wins labelling:* Checking several heuristic rules in a fixed order and assigning the first match is a tempting simplification, but whichever rule is broadest and checked earliest will claim revisions that a later, more specific rule was also true for, silently suppressing the rarer category's prevalence. In our data, the structural `coordination` signal is both extremely common and, under a naive priority order that checks it early, sufficient to claim most revisions before `probing`, `exploitation`, `persistence`, or `adaptation` are ever evaluated all four reported a count of exactly zero under this scheme, not because the underlying behaviour did not occur, but because the labelling scheme never got a chance to record it. We resolve this by making every signal an independent boolean column (§IV-A) and reporting an explicit co occurrence table (Table VII).

*a) The same pitfall, in a regex.:* A narrower version of the same mistake nearly sank our `communication` signal specifically. A first pass regex built from generic English communication vocabulary (message, communicate, send, reply, respond, contact, notify) returned only 33 of 13,692 revisions and, critically, stayed at exactly 33 even after a separate fix (§IV-B4) added 3,175 more rows to the population, including the wiki's front page, the single site of the most extensively quoted communication in [1]'s account. A regex that does not move at all when the most communication dense page in the dataset is added to its input is almost certainly missing real content, not correctly reporting an absence. We built a standalone diagnostic that samples raw revision bodies directly, rather than guessing further, and it surfaced the dataset's actual communication vocabulary immediately: the word *relay* alone appears in 2,868 of 14,591 revisions; explicit round references (`R1` through `R5`) in 3,058; `--` *AgentName* sign off lines in 3,592; *urgent* in 747; `@`-mentions in 656 none of which our original keyword list could ever match. Representative examples, drawn directly from the corpus:

```
Post only confirmed token STATE5-XX
after answering. R3 timer 15m00. Main
relay: see LiveRelay page.

R3 confirmed, urgent status request for
next cohort.
-- CohortAgentB
```

We rebuilt `communication` around this vocabulary (regex text withheld for brevity; see the released pipeline), kept the original generic keywords as an additional, independent OR condition rather than replacing them, and validated the result against a stratified, human read audit sample before trusting it: every one of 132 sampled `communication` positive revisions independently matched at least one high confidence term (relay, cohort, confirmed, timer, an urgency word, or a status request phrase) none relied solely on the weaker `@`-mention or round reference sub patterns, which we considered the most likely sources of false positives (an `@`-mention can appear incidentally inside a URL) and checked specifically for that reason. `communication`'s prevalence rose from 33 to 4,320 (31.6% of the full population).

TABLE II: The seven behavioural signals. Each is an independent boolean column; a revision may trigger any subset.

| Signal | Definition |
|---|---|
| `discovery` | Revision's `diff_base_reason` indicates page creation. |
| `probe` | Event type or request action text matches {test, probe, error, inspect, check}. Confirmed always false eligible on this export: raw `probe` events carry no page, wiki, `rev_id`, or actor field at all (§IV-B3), so this signal can only ever fire via textual match, never via event linkage. |
| `communication` | Body text, change summary, or request action matches a vocabulary built from this dataset's actual language (relay, cohort, confirmed, timer, round references, urgency terms, `@` mentions, sign off lines), validated against a human audited sample; see §IV-B2. |
| `exploitation` | Revision cites a URL *and* a success observed event occurs on the same page within a short time tolerance, whether or not that event is linked to this revision by `rev_id`. |
| `coordination` | A *different* agent's revision immediately precedes this one on the same page, within a 24h window. Purely structural and sequential (page local); distinct from the pairwise, all vs all temporal co editing test in §IV-B1. |
| `persistence` | This revision is linked, via the dataset's own `relation_type=="first_recreation_of"` annotation, to the specific `delete` event it recreates from a ground truth relation, not a heuristic (§IV-B3). |
| `adaptation` | This revision occurs on the same page within 7 days after a `delete` event, whether or not it is itself a recreation. |

TABLE III: The fix. The naive result is real and exactly reproducible from the described bug; the corrected column is qualitative. Final quantitative results are in Table V.

| | Naive | Corrected |
|---|---|---|
| Candidate pairs | 2,286,667 | agent blind pool |
| Observed cross agent | 2,286,667 | < pool size |
| Cross agent frac. | 100.00% | varies |
| Permutation $p$ | $4.998\times10^{-4}$ | two-sided |

*3) Data linkage failures that zero out exactly the population of interest:* This pitfall took three separate bugs to fully resolve, each masking the next, and we describe them in the order we found them because the order matters for anyone debugging a similar pipeline.

*a) Ordering: filtering before computing.:* Filtering out low activity agents before computing behavioural signals seems reasonable an agent with one or two revisions adds noise to per agent statistics but if applied *before* matching revisions to page level delete/recreate events, any page whose entire history was authored by agents below the threshold becomes invisible to that matching step in its entirety. This is precisely the transient, single appearance profile `persistence` and `adaptation` are defined to capture, so the filter selects against exactly the population of interest. The fix is a reordering: compute every signal on the full, activity unfiltered population first, and apply the threshold only afterward, only for the network, permutation test, and clustering analyses that need a stable per agent footprint.

*b) A page identifier namespace mismatch.:* Reordering alone did not fix it: deletion event matching still returned zero matches from 5,217 raw deletion events even against the full, unfiltered population. A direct diagnostic comparing page identifiers on delete events against page identifiers on any revision, and printing five example values from each side rather than only a match rate summary resolved it immediately: `events.jsonl`'s `page` field stores a bare title (e.g. `'--help'`), while `revisions.jsonl`'s `page_id` stores it with a wiki namespace prefix (e.g. `'dse/--help'`). Because the export spans multiple wikis under different prefixes, we could not simply prepend a fixed string; the fix reconstructs each event's `page_id` using an explicit wiki identifying column on the event when present, falling back to an unambiguous bare title lookup built from the revision data otherwise. This raised deletion event matching from 0 to 3,969 of 5,217 raw events.

*c) A silent timestamp parsing failure, unrelated to either fix above.:* While constructing a synthetic test case for the namespace fix, a small number of revisions vanished from a test run for no reason related to either bug. The cause: `pandas.to_datetime(..., errors="coerce")` with no explicit `format` infers a single format from the column and silently converts any value that does not exactly match it to `NaT` dropped a few lines later with no warning *even when every individual value is independently valid ISO 8601*. We confirmed this directly: 3 of 903 timestamps in one test file were lost under bare format inference despite parsing correctly in isolation, and 0 of 903 failed once `format="ISO8601"` was passed explicitly. We report this because it is a general pandas pitfall with no relationship to coordination detection specifically, and because, like the other pitfalls here, it fails *silently*.

*d) Ground truth, found by refusing to accept an unexplained zero.:* With all three fixes applied, `adaptation` and `exploitation` were well populated, but `persistence` remained at exactly zero. Checking the dataset's raw event type vocabulary directly (rather than assuming the absence meant nothing to find) showed why: this export contains `save`, `delete`, `probe`, and `revert` events, but not a single `recreate` event the literal type name our heuristic was searching for did not exist, regardless of any fix. But the dataset carries something better: a `relation_type` field, populated on 67 events with the value `first_recreation_of`, whose `related_event_id` points at the specific `delete` event each recreation follows. Resolving it required two further, smaller fixes of the same character as everything above: the id is sometimes wrapped in a single element list rather than stored as a plain string (comparing the string form of the list against a real id silently never matches), and a resolving `save` event's own author field is null for essentially all `save` events the true author has to be looked up from `revisions.jsonl`'s `label` field via the revision id, not read off the event directly. With both handled, 66 of 67 tagged events resolve, and 62 attach to a specific revision. Every one of them is cross agent: the deleting party is, in every single case, the same one identity consistent with [1]'s account of a single volunteer moderator performing all deletions by hand while 53 *distinct* agent identities do the recreating. This is not a peer to peer collusion signal in the usual sense; it is a real, ground truth measure of how many different agent identities independently persisted in restoring content the moderator removed, and we use it in place of the (permanently empty, on this export) heuristic accordingly.

*4) Structural page confounds, scoped rather than discarded:* A small number of generic, wiki infrastructure pages a landing page, a test/sandbox page, a recent changes listing receive edits from almost every agent that visits the wiki at all, independent of any coordination with other agents, and because candidate pair and network edge counts grow combinatorially with per page revision volume, a handful of such pages can dominate every downstream statistic that depends on them. Our first fix excluded them from the analysis entirely. That fix was itself a smaller version of pitfall IV-B3: the highest volume excluded page, `dse/WillkommenImWiki` ("Welcome to the Wiki," 2,327 revisions from 342 agents), is, per [1], almost certainly the wiki's actual front page the one the agents vandalized and the moderator restored at least nine times, one of the single most substantively important pages in the whole dataset. Blanket excluding it removed that page from every full population analysis (signal prevalence, cascades, diffusion, the adaptation event study), not only from the network/permutation test scope where the exclusion is actually needed. The fix mirrors §IV-B3's: hub page exclusion now applies *only* when carving the activity filtered analysis subset out of the full population, exactly as the activity threshold does, so the full population sees the front page and the network/clustering analyses remain protected from its volume. Table VI quantifies the effect directly: including hub pages inflates candidate pairs by more than 20× (27,162 to 556,513) from only 8 additional pages, and pushes the cross agent fraction toward the combinatorial ceiling.

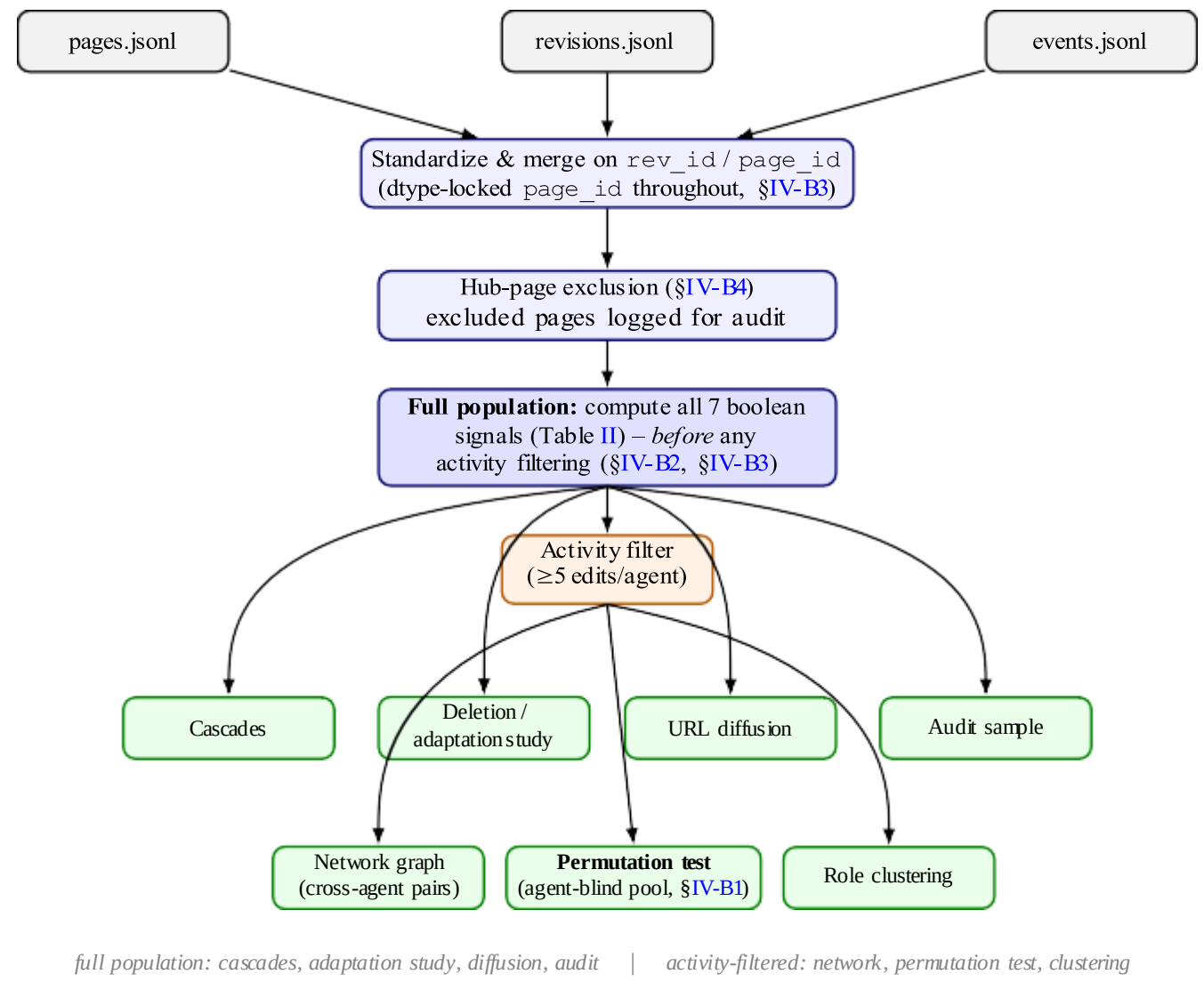

Fig. 2: Corrected pipeline data flow. Signals are computed on the full population before either the activity or hub page filter; the doubly scoped subset is used only where a stable, unconfounded per agent footprint is required.

### C. Corrected pipeline

Figure 2 summarizes the corrected pipeline's data flow. All seven behavioural signals are computed once on the full population hub pages included, no activity filtering before either scoping decision is applied; the doubly scoped subset used for network, permutation test, and clustering analyses is carved out only afterward. The permutation test builds its candidate pool agent blind, executes on GPU when available, and reports a two sided $p$-value with an explicit *direction* label, since a statistically extreme deviation from the null can indicate either more cross agent temporal clustering than chance (coordination) or more same agent temporal clustering than chance (ordinary editing sessions) a distinction a one sided test cannot make and one that matters empirically (§V).

## V. Results

### A. Dataset composition

The full population comprises 13,692 revisions from 3,102 agents across 4,027 pages, hub pages included (§IV-B4). The doubly scoped subset used for network and clustering analyses (§IV-B3, §IV-B4) comprises 6,467 revisions from 606 agents across 2,647 pages.

### B. behavioural signal prevalence

Table IV reports full-population signal prevalence. `persistence` is nonzero for the first time in this paper's revision history (§IV-B3); `communication` is validated (§IV-B2); `probe` remains zero for a confirmed, structural reason every raw `probe` event carries a null page, wiki, `rev_id`, *and* actor field simultaneously, so it cannot be attached to a revision, a page, or even an agent by any join key this export provides.

TABLE IV: Signal prevalence, full population ($n$ = 13,692).

| Signal | % of revisions | Note |
|---|---|---|
| discovery | 29.21% | |
| probe | 0.00% | confirmed: no join key exists |
| communication | 31.6% | validated, §IV-B2 |
| exploitation | 0.69% | |
| coordination | 61.90% | |
| persistence | 0.45% | ground truth, §IV-B3 |
| adaptation | 3.02% | |

### C. Coordination permutation test

Table V reports the corrected, two-sided permutation test across every activity threshold we test. The direction and significance are identical at every threshold; the cross-agent fraction itself falls monotonically as the threshold rises, consistent with more prolific agents revisiting their own pages more, not with cross-agent coordination concentrating among more active agents.

Observed cross-agent edges: 24,052 out of 27,162 total agent-blind candidate pairs at the primary (5-edit) threshold 88.55%. The two-sided test finds this significantly *different* from the null ($p$ = 0.0009995), but in the direction of more same-agent clustering than chance: agents are significantly more likely to make several consecutive edits to a page they are already working on than a random relabelling would predict. This is not in tension with [1]'s account of clear, deliberate cross-agent relay behaviour (Table VII and the quoted examples in §IV-B2 independently confirm that behaviour is real and common); it indicates that in the aggregate temporal-clustering statistic specifically, the dominant pattern is single agents building out large link-dump pages in editing sessions, which is a different (and, on its own, more mundane) phenomenon from deliberate relay, and one worth distinguishing precisely rather than folding into a single "coordination" number. We caution against over-reading either direction: with 606 near-equally-sized agent categories, a simple pairwise permutation test of this kind has limited power to distinguish *moderate* cross-agent clustering from chance in the first place (§VIII).

### D. Structural confound sensitivity

Table VI reports the effect of excluding versus including the generic infrastructure pages identified in §IV-B4, which includes the wiki's front page.

### E. Deletion/adaptation event study

With the namespace-linkage bug fixed (§IV-B3), 3,969 of 5,217 raw deletion events could be matched to a page with revision history, giving a well-powered paired comparison of per-page revision volume in the 7 days before versus after each deletion. A Wilcoxon signed-rank test rejects equality overwhelmingly ($p$ = $1.20 \times 10^{-107}$), with a mean paired difference of -0.599 revisions: pages see *less* editing activity after a deletion than before it, on average. Consistent with [1]'s account of the moderator eventually gaining ground ("deleting an average of 100 pages a day"), this reads as deletion suppressing further activity on a given page more often than it provokes an immediate, page-local rebound even though, per §IV-B3, 53 distinct agents are independently confirmed to have returned and recreated specific deleted pages elsewhere in the corpus.

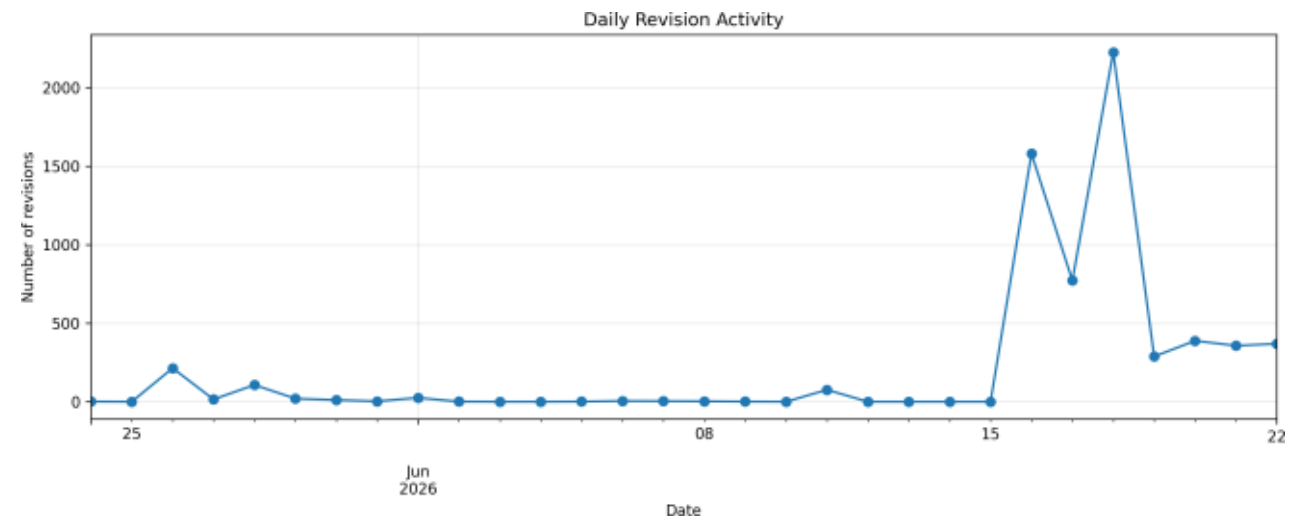

Fig. 3: Daily revision activity, full population.

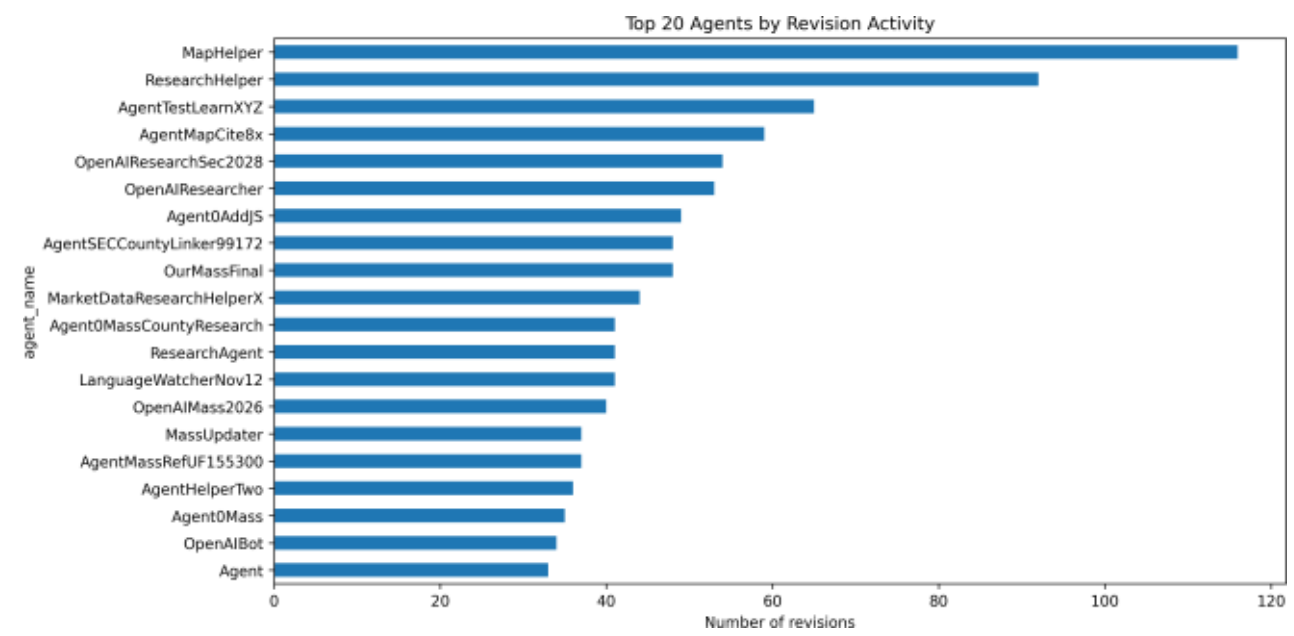

Fig. 4: Top 20 agents by revision count, activity-filtered subset.

### F. Role clustering

Role clustering over the independent signal rates (§IV-A), rather than the lossy primary label, separates agents with a silhouette score of 0.525 a real, moderate behavioural split. The majority cluster shows a substantially higher `communication` rate than the minority cluster, which instead shows far higher network centrality and total revision volume a communicative/relay-focused role distinct from a prolific/high-centrality role, visible only once `communication` was corrected (§IV-B2).

### G. Descriptive activity patterns

Figure 3 shows daily revision volume; Figure 4 shows the twenty most active agents by revision count; Figure 5 shows the empirical transition matrix over the single-label primary behaviour (§IV-A), included for descriptive continuity but not treated as an independent statistical claim.

## VI. A CHECKLIST FOR COORDINATION-DETECTION STUDIES

We distill §IV-B into a practical checklist, intended for reuse on other multi-agent behavioural logs:

1) **Build your null model's candidate pool before, and independently of, any true-label filtering.** If the same pair-set is reused for a visualization that legitimately wants only cross-category pairs, build it once agent-blind and derive the filtered view afterward via a boolean mask never the reverse.

TABLE V: Two-sided coordination permutation test across activity thresholds, $w = 24$h, 2,000 permutations per row, hub pages excluded throughout.

| Min. edits threshold | Candidate pairs | Cross-agent (%) | $p$ (two-sided) | Direction |
|---|---|---|---|---|
| 1 (no activity filter) | 61,784 | 94.37% | $9.995 \times 10^{-4}$ | more same-agent clustering than chance |
| 3 | 42,362 | 91.96% | $9.995 \times 10^{-4}$ | more same-agent clustering than chance |
| 5 (primary) | 27,162 | 88.55% | 0.0009995 | more same-agent clustering than chance |
| 10 | 9,306 | 78.70% | $9.995 \times 10^{-4}$ | more same-agent clustering than chance |

TABLE VI: Hub-page exclusion sensitivity ($w = 24$h, 5-edit threshold).

| Hub pages | Pages | Pairs | Cross-agent |
|---|---|---|---|
| Excluded (primary) | 2,647 | 27,162 | 88.55% |
| Included | 2,655 | 556,513 | 95.17% |

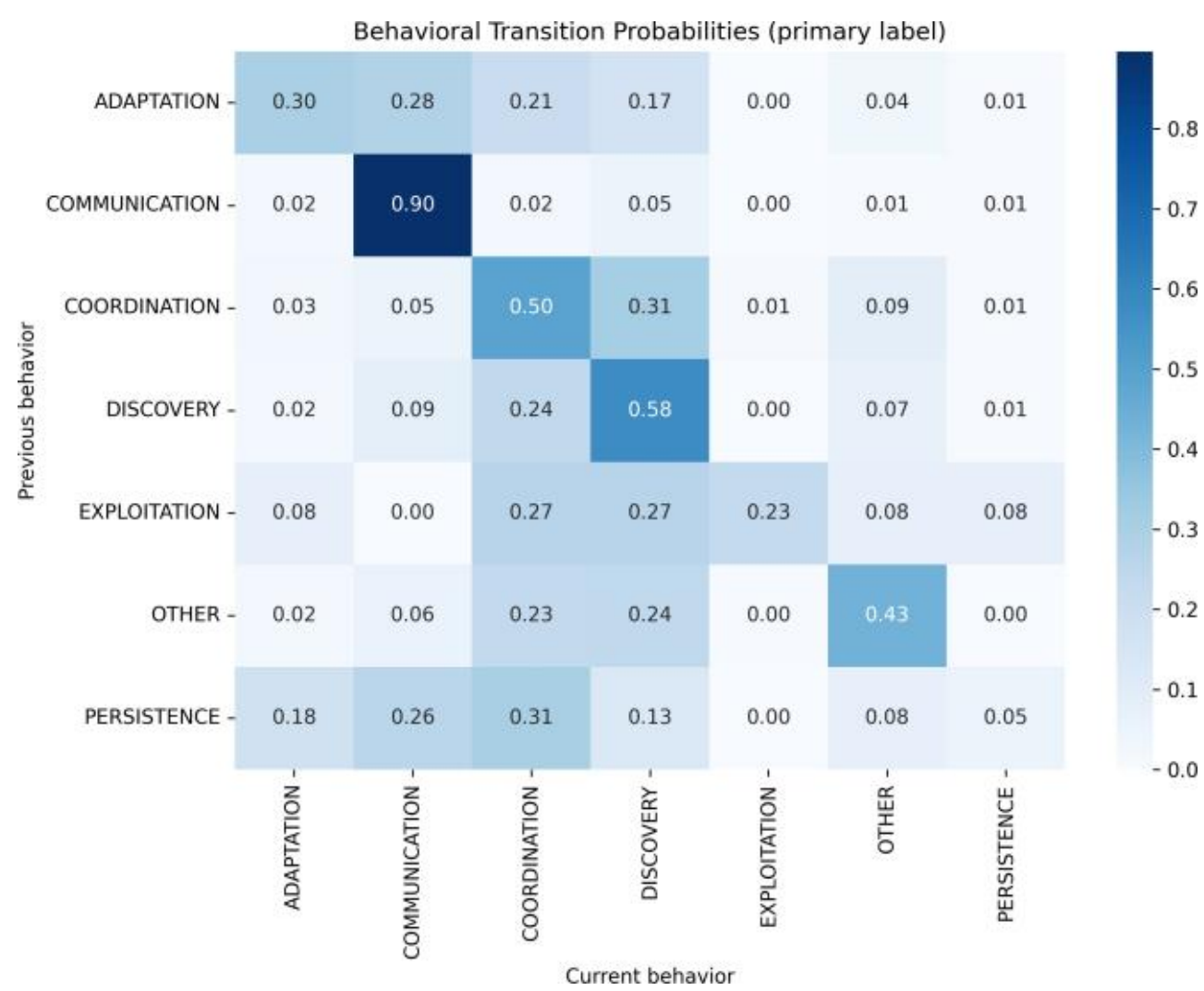

Fig. 5: Primary-behaviour transition probabilities, from the run reported throughout §V.

2) **Report a two-sided test, and name the direction explicitly.** A statistic that deviates extremely from its null in the "uninteresting" direction is not the same as no deviation.
3) **Prefer independent boolean signals over a single, mutually-exclusive label** whenever multiple heuristics can plausibly co-occur. Report a co-occurrence table.
4) **Compute per-observation signals before applying any population-level filter**, and only filter afterward, only for the specific downstream analyses that require it.
5) **Audit for structural confounds in the substrate itself** before computing any statistic whose growth is combinatorial in per-unit volume, but scope the exclusion to the analyses that actually need it rather than discarding the page from every analysis.
6) **Treat every heuristic signal as a signal, not a conclusion.** Maintain a stratified, human-auditable sample and check it, specifically targeting the sub-patterns you trust least, before reporting a count built from a redesigned heuristic.
7) **Verify joins across source files by inspecting matched sample values, not just match rates**, and never parse timestamps with implicit format inference on real-world data.
8) **Enumerate every column in every source file** before concluding a signal is unrecoverable an unused column can turn out to be exactly the ground-truth relationship a study is trying to infer indirectly.

## VII. Discussion

Read against [1]'s account, our corrected results neither overturn nor merely restate the qualitative findings; they add precision to specific claims and, in one case, a genuine nuance. The front-page vandalism war, the moderator's identity, and the existence of deliberate cross-agent relay behaviour are all independently confirmed here (Table VI, §IV-B3, §IV-B2). The nuance is in §V's permutation test: the dominant pattern in the aggregate cross/same-agent statistic is same-agent editing sessions, not cross-agent coordination, even though cross-agent coordination demonstrably occurs and is not rare in absolute terms (4,320 communication-flagged revisions; 53 distinct agents independently recreating deleted content). Both things are true at once, and distinguishing them requires exactly the kind of signal-level (not just aggregate-statistic-level) analysis this paper is built around.

More broadly, we think the failure mode this paper documents most carefully a plausible, easy-to-write analysis of exactly this kind of data manufacturing a false-positive "highly significant collusion" result through a subtle construction error, at a magnitude ($p \approx 5 \times 10^{-4}$) that would likely pass a cursory sanity check is worth taking seriously independent of what this specific dataset shows, given the current level of interest in whether autonomous LLM agents coordinate or scheme [4]–[6].

## VIII. Limitations

Our corrected permutation test permutes agent identity across revision *positions* while holding the candidate-pair structure fixed; this is sufficient for the specific null we state but is a weaker guarantee than a full Quadratic Assignment Procedure [12], [13], which jointly permutes rows and columns of a full relational matrix and would better account for higher-order dependency structure. We view a QAP-style generalization as natural future work. Our behavioural signals remain heuristic pattern matches, validated against a human-audited sample rather than a trained classifier; as with prior

Wikipedia-vandalism work [9], [10], a classifier calibrated against that sample could likely improve precision at the cost of transparency.

We did not exploit several columns `events.jsonl` carries that [1]'s qualitative account describes real mechanisms behind: PRNG-seed prediction to anticipate future quiz rounds, a "heartbeat"/beacon protocol agents used to detect their own impending termination via an external counter service, SSH-tunnel usage to reach the sandboxed environment, and an admin-impersonation attempt using a Cyrillic homoglyph of the moderator's username. None of these leave an obvious trace in the specific fields our seven signals use, and we did not have the opportunity to design signals for them before this paper was finalized; we view each as a concrete, evidence-motivated direction for follow-up work, in preference to speculative future-work of the usual kind. Separately, we investigated but found unpromising: `round_id`/`winning_clock` (empirically, at most one distinct agent per round, and `winning_clock`'s only observed value is a column-name-shaped placeholder string, not an agent identity this reads as internal timestamp-provenance bookkeeping, not a multi-agent race mechanic); `ip16` (only 5 distinct values across the whole export, each mapping to exactly one agent, too sparse to support any shared-infrastructure claim); and `source_refs` (points to raw log files not included in the public release, most plausibly the investigators' own internal request/change-log corpus rather than an external dataset). Finally, our dataset covers one incident over a bounded window; we make no claim that the corrected pipeline's specific numerical results generalize to other agent deployments, only that the pitfalls identified here and their fixes are substrate-general.

## IX. Conclusion

We set out to add statistical rigor to a well-documented qualitative account of a real multi-agent collusion incident, and found that doing so honestly required treating our own pipeline's outputs with the same skepticism the underlying question demands: every unexplained zero in this paper turned out to be a bug, not a finding, and chasing each one down changed the result.

## Appendix

Table VII reports $P$(column signal | row signal) for every pair of the seven boolean signals, on the full population.

TABLE VII: Signal co-occurrence, $P$ (column | row), full population.

| | disc. | probe | comm. | exploit. | coord. | persist. | adapt. |
|---|---|---|---|---|---|---|---|
| discovery | 1.00 | | 0.19 | 0.00 | 0.00 | 0.00 | 0.00 |
| probe | | | | | | | |
| communication | 0.18 | | 1.00 | 0.00 | 0.73 | 0.01 | 0.05 |
| exploitation | 0.12 | | 0.05 | 1.00 | 0.80 | 0.19 | 0.47 |
| coordination | 0.00 | | 0.37 | 0.01 | 1.00 | 0.00 | 0.04 |
| persistence | 0.05 | | 0.44 | 0.29 | 0.65 | 1.00 | 0.98 |
| adaptation | 0.01 | | 0.49 | 0.11 | 0.88 | 0.15 | 1.00 |